\documentclass[aps,prl,twocolumn,reprint,letterpaper]{revtex4-1}

\usepackage{graphicx}     

\newcommand{\prom}[2]{\left \langle #1 \right \rangle_{#2}}

\newcommand{\qvec}{\mathbf{q}}

\newcommand{\Eq}[1]{Eq.~(\ref{eq:#1})}
\newcommand{\iq}{{\rm if}}
\newcommand{\qv}{{\rm fv}}
\newcommand{\iqv}{{\rm iv}}

\begin{document}

\title{Premelting induced smoothening of the ice/vapor interface}





\author{Jorge Benet,$^\ddag$, Pablo Llombart, Eduardo Sanz and Luis G. MacDowell$^\dag$}
\affiliation{Departamento de Qu\'{\i}mica-F\'{\i}sica, Facultad de
Ciencias Qu\'{\i}micas, Universidad Complutense de Madrid, 28040 Madrid, Spain}
\email[\dag]{lgmac@quim.ucm.es}
\altaffiliation{$^\ddag$Department of Physics, University of Durham, South Road,
Durham DH1 3LE, United Kingdom}




\begin{abstract}
We perform computer simulations  of the quasi-liquid layer
of ice formed at the ice/vapor interface close to the 
ice Ih/liquid/vapor triple point of water.
Our study shows that the two distinct surfaces bounding the film 
behave at small wave-lengths
as 
atomically rough
and independent ice/water and water/vapor
interfaces.
For long wave-lengths, however, the 
two surfaces couple, large scale parallel fluctuations
are  inhibited and the ice/vapor interface
becomes smooth.  Our results could help explaining the complex
morphology of ice crystallites.
\end{abstract}

\maketitle


Nakaya summarized his research on snow flakes in a famous Haiku:
"they are letters sent to us from the sky" \cite{sunagawa87}. 
Indeed, the habit of ice crystals grown from bulk  vapor 
change from plates, 
to columns, to plates and yet back to columns as temperature is cooled 
down below the triple point, with the well known dendritic patterns appearing
at sufficiently high super-saturations.\cite{libbrecht05}  Accordingly, the 
final growth form of a tiny ice crystal conveys detailed information 
on the atmosphere where it grew.\cite{pruppacher10}

At a macroscopic level, it is well known that changes in ice crystal
habits result
from a crossover in the growth rates of the basal and prismatic
faces,  but exactly what structural transformations occur on
the surface to drive this crossover is far from being
understood.\cite{liu95,sunagawa87,libbrecht05,furukawa07} 
Kuroda and Lacmann explained the
crossover in crystal growth rates as a result of
the formation of a thin {\em quasi-liquid} layer on the
ice surface which could 
set up at different temperatures depending on the crystal facet.\cite{kuroda82} 

The hypothesis that ice could exhibit a quasi-liquid layer dates
back to Faraday, and the formation of such layer on solid surfaces
is now well characterized theoretically as
a premelting surface phase transition.\cite{lipowsky82}
Experimentally, the advent 
of modern optical and surface scattering techniques has allowed
to gather ample evidence as regards the existence of a premelting liquid film
on the surface of 
ice.\cite{dash06,elbaum91,lied94,dosch95,bluhm02,sazaki12,murata15} 
Unfortunately, the 
relatively high vapor pressure of ice makes it very difficult to achieve
sizable equilibrium crystals,\cite{libbrecht05} 
while the presence of impurities has
a very large impact on surface structure.\cite{wettlaufer99,bluhm02} 
Accordingly, many other relevant properties, such
as the premelting temperature, the thickness of the quasi-liquid layer or
the presence of surface melting remain a matter of debate.\cite{dash06}

One particularly important structural property with large impact on crystal
growth rates is the surface roughness.\cite{burton51,weeks79,kuroda82} Contrary
to smooth or singular facets, which have a limited number of defects and
serve as basis for most crystal growth models, 
rough surfaces present diverging height fluctuations which do
not differ macroscopically to those found in a fluid interface. As a result,
rough crystal planes with correlation lengths that are larger than the
crystallite 
disappear and become
round.\cite{jayaprakash83,rottman84,chaikin95} 
More importantly, as far as crystal habits
and growth forms are concerned, the roughening of a surface has dramatic
consequences on the dynamics, as it signals a crossover from a two dimension
nucleated growth, to a faster Knudsen mechanism that is linear in the
saturation.\cite{weeks79,kuroda82}  
Unfortunately, this phenomenology has been established only for
rather simple interfaces,\cite{weeks79} and the role of a premelting film in
the surface roughness is largely unknown. 

Here we perform  computer simulations of a premelting layer
on the primary prismatic facet of
the ice/vapor interface. Our study reveals that the structure
and fluctuations of the surfaces bounding
the quasi-liquid layer at small length-scales are very much
like those of atomically rough and independent ice/water and water/vapor interfaces. 
However, the finite equilibrium thickness of the premelting layer 
below the triple point drives the long-wavelength structure of
the interface from rough to smooth. Our results clarify why
the facets of ice crystals remain recognizable up to the triple point,
and suggest the formation of a premelting layer could
slow down growth kinetics, as required to explain ice crystal
growth habits in the atmosphere.\cite{kuroda82}

\begin{figure}[t]
\centering
\includegraphics[clip,scale=0.30]{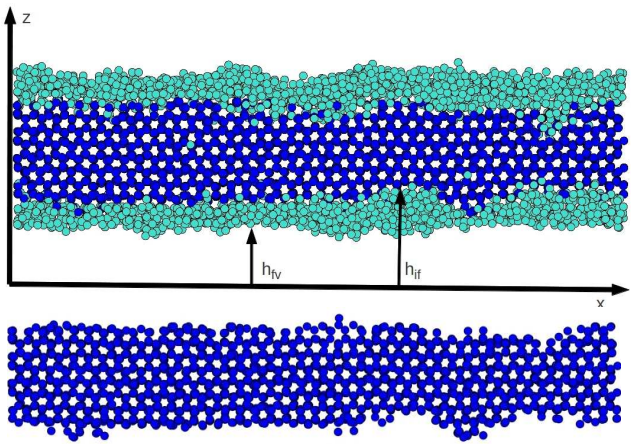}
\caption{Snapshot of the ice/vapor interface during the course
of our simulations. Top: A quasi-liquid layer of disordered molecules
is clearly seen on top of bulk ice. The order parameter allows
us to distinguish between an ice/film and a film/vapor surface.
Bottom: Same figure with liquid-like atoms removed.
\label{snapshot}
}
\end{figure}

Our study is performed with the TIP4P/2005 model of 
water,
\cite{abascal05b} 
which
has been shown to reproduce with remarkable accuracy a large number of bulk and
surface properties of (liquid) water and ice.\cite{abascal11} A slab of
equilibrated bulk ice with several thousand molecules
is placed in contact with vacuum 
inside a large ortorhombic simulation box, such that an interface
of surface area $A=L_x L_y$ is exposed parallel to the $xy$ plane. 
Surfaces thus prepared exhibit a very large heterogeneity of vacancy energies,
with a strong dependence on the proton ordering arrangements.\cite{watkins11}
For this
reason we prepare our initial samples using a special purpose Monte Carlo
algorithm that suitably samples the hydrogen bond
network.\cite{buch98,rick03,macdowell10}
Averages are then collected using Molecular Dynamics with the Gromacs package
for about half a
microsecond,\cite{supplemental13,bussi07,rommelse87,goldenfeld92,macdowell15,akutsu01,fisher83,akutsu15,benet15} well above the 
expected relaxation time for the ice/water interface.\cite{benet14}
Performing the
simulations  along the
sublimation line at a temperature $\Delta T=T-T_t$ just 2~K below the
triple point temperature, $T_t$,\cite{rozmanov11}
the first few ice layers  melt and form a premelting
quasi-liquid layer (Fig.\ref{snapshot}), 
as noticed earlier.\cite{conde08,bishop08,pan11,limmer14} 
The nature and size  of this layer may be 
quantified using the $\bar{q}_6$ order parameter,\cite{lechner08} 
which has been optimized to
discriminate ice like and water like molecules from a study of molecular
correlations up to second nearest neighbors.\cite{sanz13}  
To get rid of vapor molecules, we
identify the premelting layer as the largest cluster of water molecules, and
find an average thickness of $\ell=0.9$~nm, in reasonable agreement with experimental
observations,\cite{bluhm02,murata15} and 
recent simulations.\cite{conde08,bishop08,pan11}
Here, we attempt to characterize the quasi-liquid film
in terms of two fluctuating ice/film (\iq), and film/vapor (\qv) surfaces,
which we locate by locally averaging the heights of the outermost solid and liquid molecules of the layer,
respectively (Fig.\ref{snapshot}). 
Comparing our results for the ice/vapor interface with our previous study of 
the ice/water interface will prove insightful.\cite{benet14c}  
Since we aim at
studying large wavelength fluctuations, we prepare the exposed faces with
an elongated geometry, with box side $L_x\gg L_y$.  This allows us to identify 
ice/film, $h_{\iq}(x)$ and film/vapor $h_{\qv}(x)$ surface profiles along the 
largest axis, $x$. These film profiles are then Fourier transformed to yield
the spectrum of surface fluctuations.\cite{supplemental13}

For the purpose of studying the fluctuations of the quasi-liquid layer,
it is convenient to define the quantity
$\Gamma_{\alpha\beta}(q_x)$, in terms of the thermal
averages of Fourier amplitudes $h_{\alpha\beta}$, as:
\begin{equation}\label{eq:defGamma}
  \Gamma_{\alpha\beta}(q_x) = 
  \frac{k_BT}{A\prom{ h_{\alpha\beta}(q_x) h_{\alpha\beta}^*(q_x)}{}q_x^2}
\end{equation} 
where $k_B$ is Boltzmann's constant, $q_x=2\pi n/L_x$, and
n is a positive integer. According to Capillary Wave Theory,\cite{safran94}
for a rough interface between bulk phases $\alpha$ and $\beta$, the function
$\Gamma_{\alpha\beta}(q_x)$ may
be identified with a wave-vector dependent stiffness, 
$\tilde \gamma_{\alpha\beta}(q_x)$
whose $q_x\to 0$ limit is the macroscopic stiffness of the
interface,\cite{hoyt01,davidchack06,hartel12}
corresponding exactly to the surface tension for fluid/fluid
interfaces.\cite{mueller96,chacon05}
In the forthcoming
exposition we concentrate on the primary prismatic plane (pI) at $\Delta T=-2$~K 
and study
the fluctuations propagated along  the basal [Basal] and secondary prismatic
[pII] directions. 


\begin{figure*}[t]
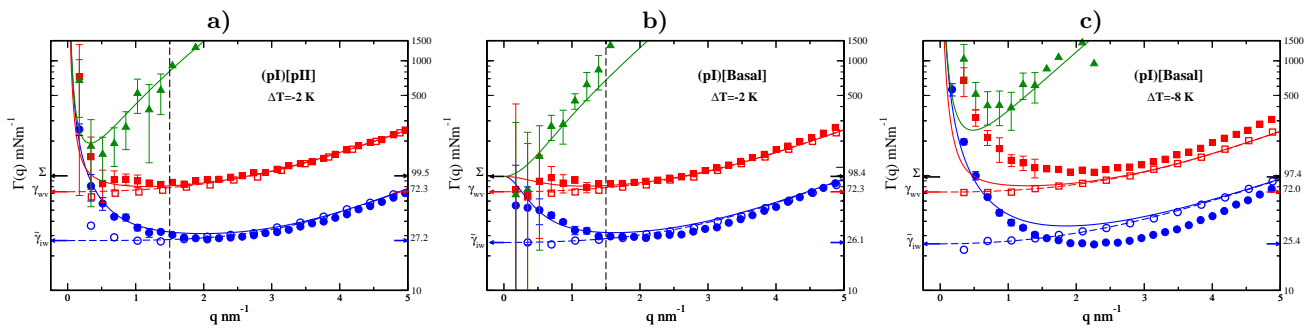

 \begin{tabular}{ccc}
 {\bf a)} & {\bf b)} & {\bf c)} \\
 \includegraphics[clip,scale=0.22]{gamma.pi.pii.248k.eps} &
 \includegraphics[clip,scale=0.22]{gamma.pi.basal.248k.eps} & 
 \includegraphics[clip,scale=0.22]{gamma.pi.basal.242k.eps} 
 \end{tabular}
\caption{
Fluctuations of the premelting film on the primary prismatic plane.
The plot displays effective wave-vector dependent stiffness, $\Gamma(q_x)$ 
in log scale for (pI)[pII] at $\Delta T=-2$~K ({\bf a}); and (pI)[Basal] at
$\Delta T=-2$~K ({\bf b}) and $\Delta T=-8$~K ({\bf c}).  Results
for the  the quasi-liquid layer are shown with filled symbols,
with $\Gamma_{\iq}(q_x)$ for the ice/film (blue circles) and $\Gamma_{\qv}(q_x)$
for the film/vapor surfaces (red squares). The open symbols are
results for the ice/water (blue circles) and water/vapor (red squares)
interfaces, which are fitted to $\Gamma(q_x)=\gamma + \kappa
q_x^2 + \epsilon q_x^4$ (dashed lines) for the purpose of extrapolation
(c.f. Ref.\onlinecite{benet14c} and \onlinecite{supplemental13}). The
colored arrows indicate extrapolation to $q_x=0$, which provides the ice/water
stiffness $\tilde\gamma_{iw}$ and
the water/vapor surface tension, $\gamma_{wv}$, respectively. The black
arrow points to $\Sigma=\tilde\gamma_{iw}+\gamma_{wv}$, where the
effective stiffness of the quasi-liquid film would converge were the interface
rough.
The dashed vertical line indicates approximately the regime of $q_x$ where
the quasi-liquid surfaces cease to behave independently.
The green triangles indicate results for the coupled fluctuations
of the ice/film and film/vapor surfaces, $\Gamma_{\iqv}(q_x)$. 
Full lines are results from a fit to the model of \Eq{h2surf}.
\label{gamma}
}
\end{figure*}

The results, $\Gamma_{\iq}(q_x)$ and $\Gamma_{\qv}(q_x)$ 
obtained for
the ice/film and film/vapor surface fluctuations of the premelting
layer on the primary prismatic plane, 
either along [Basal] or [pII]
orientations
agree very nicely with those obtained for the 
corresponding  
ice/water,
$\Gamma_{\rm iw}(q_x)$, and water/vapor, $\Gamma_{\rm wv}(q_x)$,
interfaces down to $q_x^*~1.5$~nm$^{-1}$ (Fig.\ref{gamma}.a-b). This implies 
that 
for a 
quasi-liquid
layer hardly one nanometer thick, the ice/film and film/vapor surfaces
at this length-scales fluctuate
independently, with
fluctuations that can hardly be distinguished from those found at
the rough interfaces of bulk water. 
Interestingly, at $q_x\approx q_x^{*}$,
$\Gamma_{\rm iw}(q_x)$ and $\Gamma_{\rm wv}(q_x)$ are
already close to their
$q_x\to 0$ limit, and are therefore close to the corresponding macroscopic
stiffness coefficients. 

\begin{figure}[t]
\centering
\includegraphics[clip,scale=0.30]{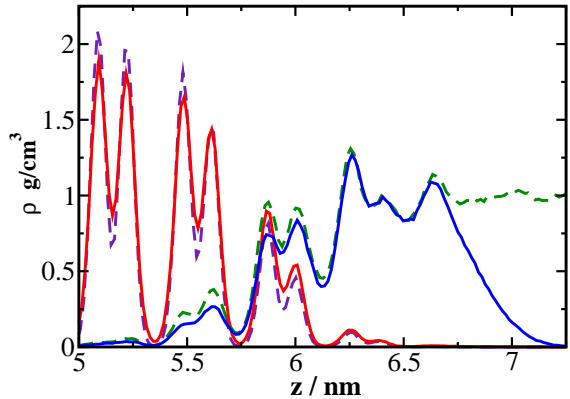}
\caption{Structure of the
ice/vapor (full lines) and ice/liquid (dashed lines) interfaces
of the primary prismatic plane
along the perpendicular direction.  The density
of solid-like molecules is shown in full-red for the
solid/vapor interface and in dashed-indigo for the solid/liquid interface.
The density of liquid-like molecules is shown in full-blue for the
solid/vapor interface and in dashed-green for the solid/liquid interface.
\label{profiles}
}
\end{figure}

The striking resemblance between the surface fluctuations of
the quasi-liquid layer and bulk water for $q_x>q_x^*$ can be understood in terms 
of 
the density profiles shown in Fig.\ref{profiles} for the primary prismatic
plane (with similar results found for the basal and secondary
prismatic planes, c.f. Ref.\onlinecite{supplemental13}).  Indeed, the density profile
of solid-like molecules from the ice/vapor interface (full-red) almost matches
that observed at the ice/water interface (dashed-indigo). Similarly,
the profile of liquid-like
molecules of the quasi-liquid layer at the ice/vapor interface (full-blue) is
very similar to that at the ice/water interface (dashed-green)
 until the very end of
the premelting film, where it obviously drops to the values
expected for the bulk vapor density. 

Below  $q_x^*$, the fluctuating surfaces start noticing
the  finite thickness of the quasi-liquid layer,  as implied by the  departure of
 $\Gamma_{\iq}(q_x)$ and $\Gamma_{\qv}(q_x)$
from the ice/water and water/vapor behavior. 
For the fluctuations in the (pI)[pII] direction, 
a sharp rise of  $\Gamma(q_x)$ for both the
ice/film and film/vapor surfaces suggest a divergence as $q_x\to 0$, and
 indicate the onset of a completely different regime,
with finite correlations at infinite wavelengths and an
effective infinite stiffness coefficient (Fig.\ref{gamma}.a). For the fluctuations
in the (pI)[Basal] direction, on the contrary, $\Gamma(q_x)$ rises above
the values expected for the ice/water and water/vapor interfaces, but
seems to attain a finite asymptotic limit for $q_x\to 0$ (Fig.\ref{gamma}.b).
These conflicting results for the (pI) interface at $\Delta T=-2$~K indicate
the proximity of a {\em roughening transition}, where the
interface depins from the underlying bulk solid.
Roughening is a Kosterlitz-Thoules transition of infinite order.\cite{chaikin95}
Not unexpectedly, the error bars observed for $\Gamma(q_x)$ are extremely large,
and  sub-averages may be collected which appear consistent with either a
rough or a smooth interface.
By simulating the same interfaces just 6~K below, we find that $\Gamma(q_x)$
also becomes divergent for the (pI)[Basal]  direction,
confirming the smoothening of the
interface just a few Kelvin below the triple point 
(as expected the divergence remains for (pI)[pII] direction, c.f.
Ref.\onlinecite{supplemental13}). 
This result is consistent with the rounding of
edges between prismatic facets observed in simulations of ice
micro-crytallites.\cite{pan11} 

The observation of a roughening transition at about $\Delta T=-2$~K
 is somewhat puzzling. Given the similarity between the structure
of the quasi-liquid layer and bulk water at short length scales, why are the long wavelength fluctuations 
so different?
Here we show how the finite thickness of the premelting film can
change completely the low wave-vector response of the ice/film and
film/vapor surfaces even under the assumption that the
corresponding stiffness coefficients are exactly  those of rough ice/water and
water/vapor interfaces, respectively.

To see this, we consider the sine-Gordon model of the solid/liquid 
interface,\cite{chaikin95,safran94}
and assume that the free energy of the ice/film layer is given solely in
terms of parameters akin to the ice/water interface:
\begin{equation}\label{eq:Hsl}
 H_{\iq} =  \int d{\bf x} \left ( \frac{1}{2}
\tilde\gamma_{iw} (\nabla h_{\iq})^2 -u \cos(k_z h_{\iq}) \right )
\end{equation}
where $\tilde\gamma_{iw}$ is the interface stiffness, $k_z=\frac{2\pi}{b}$
and $b$ is the inter-plane spacing.
In this model, the square gradient term penalizes the departure of the 
ice/film layer from planarity, while the cosine term favors by an
amount $u$ those configurations
where $h_{\iq}({\bf x})$ is a multiple of the lattice spacing.
For the film/vapor surface, we consider that the free energy is described by 
capillary wave theory, with departures from planarity penalized by the
water/vapor surface tension, $\gamma_{wv}$:\cite{nelson04} 
\begin{equation}
 H_{\qv} = \int d{\bf x} \left ( 
\frac{1}{2} \gamma_{wv} (\nabla h_{\qv})^2 + g(\Delta h)
 \right )
\end{equation} 
For an inert substrate, $g(\Delta h)$ is the interface potential,
which dictates the free energy of a planar premelting
film  of height $\Delta h$.\cite{dietrich88}
In our model, it plays the crucial role of coupling the
film/vapor fluctuations to the ice/film surface, since $\Delta h=h_{\qv}-h_{\iq}$.

To solve for this coupled Capillary
Wave + sine-Gordon model approximately, we extend a variational
theory for the sine-Gordon model due to Safran.\cite{safran94} 
The solution yields the Fourier modes of surface fluctuations,
as follows:\cite{supplemental13}

\begin{equation}\label{eq:h2surf}
\begin{array}{ccc}
 \langle |h_{\iq}^2({q_x})| \rangle & = &
 \displaystyle{     \frac{k_BT}{A} \frac{g'' + \gamma_{wv} q_x^2}{
w g'' + (g''\Sigma + w \gamma_{wv})q_x^2 + \gamma^2 q_x^4
} } \\
 & & \\
 \langle |h_{\qv}^2({q_x})| \rangle & = &
  \displaystyle{      \frac{k_BT}{A} \frac{w + g'' + \tilde\gamma_{iw} q_x^2}
{
w g'' + (g''\Sigma + w \gamma_{wv})q_x^2 + \gamma^2 q_x^4
}  } \\
& & \\
 \langle h_{\iq}({q_x}) h_{\qv}^*({q_x}) \rangle & = &
   \displaystyle{      \frac{k_BT}{A} \frac{g''}
{
w g'' + (g''\Sigma + w \gamma_{wv})q_x^2 + \gamma^2 q_x^4
} 
 }
\end{array}
\end{equation}
where $g''$ is the second derivative of the interface potential with respect
to the layer thickness, $\Sigma=\tilde\gamma_{iw} + \gamma_{wv}$, $\gamma^2 =
\tilde\gamma_{iw}\gamma_{wv}$, while
$w$ is a roughness parameter that needs to be solved self-consistently:
\begin{equation}\label{eq:selfc}
  w = u k_z^2 e^{-\frac{1}{2} k_z^2 \sum_{\qvec} \langle |h_{\iq}^2({\bf q})|\rangle }
\end{equation} 
Notice that the sum over wave-vectors confers to $w$ a dependence on the
surface geometry.\cite{supplemental13}

The above result nicely rationalizes our observations. 
At
large wave-vectors, $q\to\infty$, the system is 
{\em atomically rough}, i.e.
$\Gamma_{\iq}(q_x)\to\tilde\gamma_{iw}$ and 
$\Gamma_{\qv}(q_x)\to\gamma_{wv}$, whence 
the ice/film and film/vapor surfaces behave as rough ice/water and water/vapor
interfaces as observed in Fig.\ref{gamma} (a and b). 
 
Furthermore, the fluctuations  are then fully
independent. This can be seen by considering the cross correlations,
$\langle h_{\iq}({q_x}) h_{\qv}^*({q_x}) \rangle$, which
in this limit fall to zero. Defining the related function
$\Gamma_{\iqv}(q_x) = k_B T / A \langle h_{\iq}({q_x}) h_{\qv}^*({q_x}) \rangle
q_x^2$, consistent with \Eq{defGamma}, we find indeed that our simulation
results for
 $\Gamma_{\iqv}(q_x)$ diverge at large q (see Fig.\ref{gamma}).
This regime of large wave-vectors is consistent with observations by Limmer and Chandler, who
measured a stiffness coefficient from the ice/film fluctuations in
reasonable agreement with results for the ice/liquid interface in their
simulations.\cite{limmer14}
In the limit of small wave-vectors, $q_x\to 0$, however, we
get qualitatively different behaviors depending on the roughness parameter
(c.f. Fig.6 Ref.\onlinecite{supplemental13}).
On the one hand, 
if $w= 0$, the fluctuations diverge,
and both surfaces behave as rough interfaces with stiffness
$\Sigma$, (i.e.
$\Gamma_{\iq}(0)=\Gamma_{\qv}(0)=\Gamma_{\iqv}(0)=\Sigma$).
On the other hand, if $w\ne 0$, the fluctuations remain finite as 
$q_x\to 0$, whence $\Gamma_{\alpha\beta}(q_x)$ diverges as $q_x^{-2}$, 
indicating a smooth interface. 
Despite the atomic roughness at small length-scales, the smoothening
of the surface has dramatic consequences, since both
the crystal shape and crystal growth rate is dictated by 
$\Gamma_{\iq}({\bf q}\to 0)$.\cite{weeks79,jayaprakash83,rottman84,safran94}


In our simulations at $\Delta T=-2$~K, we observe, for the (pI)[Basal]
direction, a behavior consistent with $w=0$, corresponding to
a rough interface (Fig.\ref{gamma}.b). For (pI)[pII] direction, on  the contrary, we clearly observe  
for the smallest wave-vector accessible that $\Gamma_{\qv}(q_x)$ has largely 
exceeded $\Sigma$,
while $\Gamma_{\iqv}(q_x)$ attains a minimum well above $\Sigma$,
and then exhibits a strong divergence, 
as predicted by our model for
a smooth interface (Fig.\ref{gamma}.a).  This `roughness anisotropy' 
is consistent with \Eq{selfc}, which indicates that the roughening
temperature for our (pI)[pII] system could be higher than that
of the (pI)[Basal] system by a factor
$\approx$~1.01 given by the ratio of the ice/vapor stiffness coefficients
($\Sigma$).\cite{supplemental13}

The qualitative statements that result from our model may be made quantitative 
and extended to large wave-vectors provided
we replace $\tilde\gamma_{iw}$ and $\gamma_{wv}$ in \Eq{h2surf} 
by the phenomenological
wavevector dependent coefficients  $\tilde\gamma_{iw}(q_x)$, and
$\gamma_{wv}(q_x)$ obtained from the simulations of the ice/water and water/vapor
interfaces. 
A least square fit to the Fourier amplitudes of (pI) at 
$\Delta T=-2$~K, yields $g''\approx 8\cdot 10^{15}$~J/m$^4$ for both directions,
while the roughness parameter is $w=0$ for the (pI)[Basal] direction and $w=3.3\cdot 10^{15}$~J/m$^4$ 
for the (pI)[pII] direction, indicative of the proximity of a roughening transition.
At $\Delta T=-8$~K, the fit for both directions is consistent
with $g''\approx 12\cdot 10^{15}$~J/m$^4$ and $w\approx 8\cdot 10^{15}$~J/m$^4$,
corresponding to a smooth interface.

But how can the surface of the ice/film interface  become
smooth for small wave-vectors while the small wavelength
 structure 
remains essentially equal to that of a film of infinite depth? 
This question
can be answered by solving for the self-consistent condition,
\Eq{selfc}, with the help of \Eq{h2surf}. The result gives $w$ as the root
of a transcendental equation.\cite{supplemental13} 
For a film of infinite height, with $g''=0$,
we obtain:
\begin{equation}
 w \propto \left ( 1 + \frac{\tilde\gamma_{iw}}{w}\, q_{max}^2 \right )^{-\tau_{iw}}
\end{equation} 
where $q_{max}$ is an upper wave-vector cutoff for the fluctuations.
The above result corresponds to
the approximate solution of the sine-Gordon 
model due to Safran.\cite{safran94} The resulting transcendental equation
depends essentially  on one parameter 
$\tau_{iw}=\frac{k_BT k_z^2}{8\pi\tilde\gamma_{iw}}$. For $\tau_{iw}>1$, the root is 
$w=0$, and the surface is rough,
while for $\tau_{iw}<1$, the root is finite, and the surface is smooth.
For films of finite depth,  $g''>0$, and the situation changes.
The roots are still governed by an equation similar to the above result:
\begin{equation}
 w \propto \left ( 1 + \frac{\Sigma}{w}\, q_{max}^2 \right )^{-\mu_{\iq}}
\end{equation} 
but now the exponent is $\mu_{\iq}=\frac{\tilde\gamma_{iw}}{\Sigma} \cdot \tau_{iw}$,
which is always smaller than $\tau_{iw}$.\cite{supplemental13} Whence, 
for a rough ice/water surface with $\tau_{iw}$ close but
greater than unity,   $\mu_{\iq}$ will be well smaller than unity,
and the corresponding ice/film surface
will become smooth, even though the ice/water surface is rough.
Surprisingly, the exponent dictating the transition does
not depend on the thickness of the layer, as long as $g''$ is finite.
Only the precise value of $w$ is dictated by the premelting
thickness.\cite{supplemental13}

Our theoretical approach explains our simulation results and
is consistent with experimental observations. 
The roughening transition of the prismatic plane
has been measured for ice crystals in water\cite{maruyama05}
and vapor\cite{elbaum91,asakawa15}. 
It is found that 
from $\Delta T\approx -16$~K up to the triple point,
the ice/water interface is rough, while, 
due to the limited width of the quasi-liquid layer,
the ice/film surface remains smooth up to  about $\Delta T\approx-4$ to $-2$~K,
as suggested in our simulations.
In fact, in the atmosphere
ice crystals exhibit faceted  prismatic faces up to 0~C, even at
very low saturation.\cite{libbrecht05}
Since smooth surfaces have a slow activated
dynamics,  our results suggest it is the formation
of the quasi-liquid layer what could  actually slow down the crystal growth
rates and provide a mechanism for the change of crystal habits,
as hypothesized by Kuroda and Lacmann \cite{kuroda82}.

In summary, we have shown that close to the triple point  a quasi-liquid layer of premelting
ice  on the primary prismatic face behaves as two independent ice/water and water/vapor surfaces
at small wavelengths, but   
becomes smooth at long wavelengths. Our
results may help rationalize the role of the premelting layer
in the morphology of ice crystals.

\begin{acknowledgments}

E. Sanz and J. Benet acknowledge financial support from the EU grant
322326-COSAAC-FP7-PEOPLE-2012-CIG
and from a Spanish grant Ramon y Cajal. L.G. MacDowell 
acknowledges financial support from project MAT-2014-59678-R
(Ministerio de Economia y Competitividad).

J.B. and P.L. contributed equally to this work.

\end{acknowledgments}


%

\clearpage

\end{document}


\addtolength{\topmargin}{0.125in}

\addtolength{\oddsidemargin}{0.00in}



\onecolumngrid
\begin{center}

{\bf \large 
 Premelting induced smoothening of the ice/vapor interface
\\
Supplementary material}

\end{center}

\vspace{-0.25in}


\onecolumngrid
\begin{center}

Jorge~Benet,          
Pablo Llombart,
Eduardo~Sanz and
Luis G.~MacDowell 
\end{center}


\vspace{-0.25in}

\onecolumngrid
\begin{center}

{\it $^1$Dep. Qu\'imica F\'isica, Fac. Qu\'imica,
  Universidad Complutense de Madrid, 28040, Spain}
\end{center}


\onecolumngrid
\vspace{0.0in}
\vskip 10pt

{\samepage
{\bf
\begin{center}
Abstract
\end{center}
}
\begin{center}
\begin{minipage}{.8\textwidth}
{\small  
This file contains supplementary
information on simulations details,
additional results for (pI)[pII], (Basal)[pII]
and (pII)[Basal] systems and the 
solution for the coupled Capillary Wave + sine-Gordon model
described in the paper.
}
\end{minipage}
\end{center}
}


\section{Methods and simulation}

\subsection{Simulations}

Our study is performed with the TIP4P/2005 model of water \cite{abascal05b}.
A slab of bulk ice is first prepared from scratch using the algorithm of
Buch et al.\cite{buch98} to sample a hydrogen bond network consistent with the ice rules.
This initial configuration is then equilibrated with a Monte Carlo simulation
that samples over closed hydrogen bond loops
using a cluster algorithm by Rick and Haymet.\cite{rick03,macdowell10} It then undergoes an isotropic NpT
simulation in order to obtain the equilibrium lattice parameters at coexistence.
The equilibrated ice sample
is placed in contact with vacuum 
inside a large ortorhombic simulation box, such that an interface
of surface area $A=L_x L_y$ is exposed parallel to the $xy$ plane.
For production runs,
we performed NVT Molecular Dynamics
simulations with the Gromacs package at temperatures $\Delta T=T-T_t$ of
-2 and -8~K, with $T_t=250$~K the estimated triple point temperature of the
model.\cite{rozmanov11}
We use the Bussi-Donadio-Parrinello thermostat.\cite{bussi07} 
Lennard-Jones interactions are truncated at $R_c=0.85$~nm, and
electrostatic interactions are evaluated using Ewald summations.

The surface waves on a crystal depend on the crystal plane
that is exposed, as well as on the direction of propagation. 
We use the conventional terminology ({\bf u})[{\bf n}] to designate such
fluctuations,\cite{davidchack06,benet14} where {\bf u} is a vector perpendicular to the exposed
plane, while {\bf n} is a vector perpendicular
to {\bf u} and the direction of propagation.  The
vectors {\bf u} and {\bf n} are designated by the Miller-Bravais indices
\{h,k,l,i\}
of the hexagonal symmetry group corresponding to ordinary ice (Ih).
For the sake of brevity, we designate
sets of equivalent indices corresponding to a given plane
or direction as `Basal', for \{$0001$\}, `pI' for \{$10\bar{1}0$\} 
and `pII' for \{$2\bar{1}\bar{1}0$\}.

In order to study fluctuations with the largest possible wave-length
we prepare the exposed faces with
an elongated geometry, with box side $L_x>L_y$ (see Table \ref{tab:sim} for a description
of system size and simulation conditions).
This allows us to identify 
ice/film, $h_{\iq}(x)$ and film/vapor $h_{\qv}(x)$ surface profiles along the 
largest axis, $x$ (see Fig.1 of paper). Accordingly, for fluctuations
designated as (hkli)[h'k'l'i'], the plane (hkli) of the interface is
perpendicular to the z axis, while the direction [h'k'l'i'] is parallel
to the y axis. 
For ice/vapor systems, we considered a very large lateral
size $L_x\approx 36$~nm. On the other hand ice/water interfaces were studied
using $L_x\approx 18$~nm.

We choose the crystal orientation such that the primary prismatic
plane is exposed (pI). Configurations were prepared with the long
direction along either the basal [Basal] or secondary prismatic orientations
[pII]. This allows us to measure long wavelength fluctuations, which
are denoted as (pI)[Basal] and (pI)[pII], for fluctuations on the (pI) plane
along the [Basal] and [pII] directions, respectively.
 Less detailed results
for (Basal)[pII] and (pII)[Basal] geometries with $L_x\approx 18$~nm are also reported in
this document (see below).

To simulate ice/water interfaces, we use the configurations prepared for
the ice/vapor interface. We then equilibrate a slab of liquid water with
equal lateral dimensions than the solid. Both phases are brought 
together, and molecules of the liquid phase less than one molecular diameter
apart are removed. The compound system is then equilibrated in an Np$_z$T
simulation with the barostat along the perpendicular direction to the interface
only.\cite{benet14,benet14c} For the liquid-vapor interface, it suffices
to equilibrate a liquid slab under periodic boundary conditions. The slab
is then placed in vacuum and a liquid-vapor interface equilibrates gradually.

\subsection{Location of the interface}

 Liquid like and solid like molecules are distinguished using
the $\bar{q}_6$ order parameter by  Lechner and
Dellago \cite{lechner08}, which has been optimized to
discriminate ice like and water like molecules from a study of molecular
correlations up to second nearest neighbors. For the TIP4/2005 model
in the range of temperatures studied we chose a threshold value
of $\bar{q}_6=0.347$ to discriminate between liquid like and solid like
molecules.   

To define the film/vapor surface, we define a grid on the $xy$ plane.
For each bin on the grid, the position of the interface is determined
as an average of the 4 outermost liquid-like molecules contained within
a distance $x=1/2 \cdot \Delta_x$ and $y=1/2 \cdot \Delta_y$, with
$\Delta_x=0.91$~nm and
$\Delta_y= 1/2\cdot L_y$. A similar procedure is used to evaluate the ice/film surface,
with the position of the surface determined from the outermost solid-like
molecules. The corresponding surfaces are averaged along the $y$ direction
to obtain $h_{\iq}(x)$ and $h_{\qv}(x)$.
These film profiles are then Fourier transformed to yield
the spectrum of surface fluctuations. 

The spectrum of the ice/film fluctuations is compared in Fig.2 of the paper with
corresponding fluctuations for the ice/liquid interface at the same temperature.
For $\Delta T=-2$~K, we employed results obtained previously by
ourselves.\cite{benet14c} For  $\Delta T=-8$~K, we performed new simulations
(see Table \ref{tab:sim}).
Having equilibrated the ice/water interface as explained above, 
the spectrum of fluctuations is calculated exactly as described for
the ice/film surface.  The solid and liquid bulk
phases that were merged to prepare the interface were previously equilibrated
at the coexistence pressure of  $p=882$~bar.
Similarly, the spectrum for the water/vapor interface
is calculated as  for the film/vapor interface.

During the course of the simulations,
the center of the solid slab fluctuates and blurs the density profile. To avoid
this, we collect density profiles for short periods of 37.5~ns. This also helps
eliminate capillary wave roughening. Further details of the procedure
may be found in Ref.\cite{benet14,benet14c}.

\begin{table}
\begin{tabular}{ccccccccc}
Interphase & plane & direction & T/K & t/ns & n & $L_x$/nm & $L_y$/nm & $L_z$/nm \\
\hline
i/v & (pI)  &   [Basal]   &  248.6 &  624  &  8328 & 36.1163 &  2.2062 & 19.0000 \\
i/v & (pI)  &   [Basal]   &  242.7 &  517  &  6909 & 36.1013 & 2.2052  & 19.0000 \\
i/v & (pI)  &   [pII]     &  248.9 &  483  &  5700 & 36.7309 & 1.8034  & 19.0000 \\ 
i/v & (pI)  &   [pII]     &  242.4 &  429  &  5600 & 36.7190 & 1.8028 & 19.0000 \\
i/w & (pI)  &   [Basal]   &  242.2 &  563  &  7500 & 18.0009 & 2.1991 & 8.2517 \\
w/v & -     & -           &  242.1 &  431  &  5400 & 18.0059 & 2.2063 & 12.0000 \\
i/w & (pI)  &   [pII]     &  242.3 &  529  & 7060 &  18.3087 & 1.7978 & 8.3431 \\
i/v & (Basal)  &  [pII]     &  248.8 &  1268 & 10870 &  18.7696 & 1.8039 & 9.3319 \\
i/w & (Basal)  &  [pII]     &  248.1 & 715   & 9533 &  18.7696 & 1.8039 & 9.3319 \\
i/v & (pII)  &  [Basal]     &  248.3 & 257  & 3426 &  18.0577 & 2.2045 & 9.5000 \\
i/w & (pII)  &  [Basal]     &  248.7 & 603   & 8036 &  18.0134 & 2.1991 & 8.0808 \\
\hline
\end{tabular}
\caption{\label{tab:sim}
Table with detailed description of the simulations performed in this work. For
each system studied we describe the interface arrangement,
temperature, $T$, length of simulations, $t$
and size of simulation box $L_x$, $L_y$, $L_z$. All
simulations are performed at coexistence.
}
\end{table}

\subsection{Summary of results for the (pI) plane.}

\begin{figure}[t]
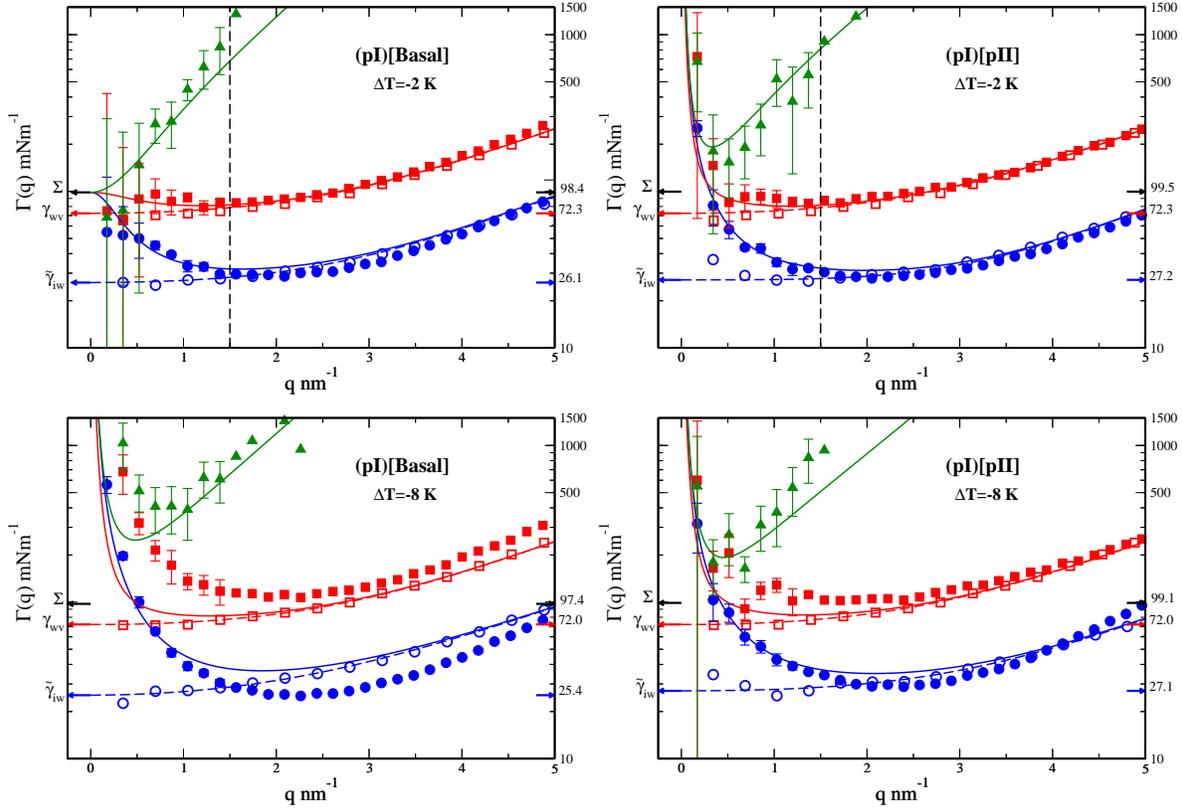

\centering
\begin{tabular}{cc}
\includegraphics[clip,scale=0.30]{gamma.pi.basal.248k.eps} &
\includegraphics[clip,scale=0.30]{gamma.pi.pii.248k.eps} \\
\includegraphics[clip,scale=0.30]{gamma.pi.basal.242k.eps} &
\includegraphics[clip,scale=0.30]{gamma.pi.pii.242k.eps}
\end{tabular}
\caption{
Fluctuations of the premelting film for the primary prismatic plane
at $\Delta T=-2$~K (top) and $\Delta T=-8$~K (bottom).
 The plot displays wave-vector dependent stiffness, $\Gamma(q)$ in log scale.
Left column corresponds to (pI)[Basal], right column to (pI)[pII].
\label{t242}
}
\end{figure}

Here we collect all results of $\Gamma(q)$ for the (pI) plane.
At $\Delta T=-2$~K, (pI)[Basal] appears rough, and (pI)[pII] appears
smooth. At $\Delta T=-8$~K, both sets of fluctuations are
consistent with a smooth interface, as can be seen from the divergence of
$\Gamma(q)$ in the limit $q\to 0$  (Fig.\ref{t242}). 
For  $\Delta T=-8$~K,
however, the coexistence pressure of the ice/water interface is now about 900 MPa,
and the short wave-length fluctuations of the ice/vapor interface differ
considerably from the ice/liquid and liquid/vapor results.
Note that the presence of atomic roughness at a small lengthscale within a
smooth surface is consistent with a pre-roughening scenario suggested by
Rommelse and den Nijs.\cite{rommelse87}

\subsection{Results for the basal and secondary prismatic planes}

\begin{figure}[t]
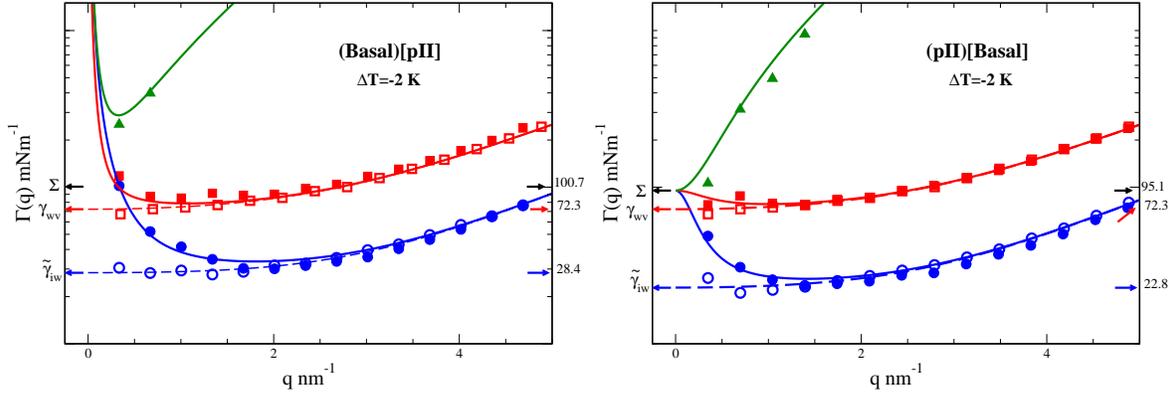

\centering
\includegraphics[clip,scale=0.30]{gamma.basal.pii.248k.eps}
\includegraphics[clip,scale=0.30]{gamma.pii.basal.248k.eps}
\caption{
Fluctuations of the premelting film for basal and secondary prismatic plane.
The plot displays $\Gamma(q)$ in log scale for (Basal)[pII] (left)
and (pII)[Basal] arrangements, respectively, for temperature $\Delta T=-2$~K.
All symbols as in Fig.2 of the paper.
\label{bypii}
}
\end{figure}

We have also studied the fluctuations on the (Basal) and (pII) planes
for $\Delta T=-2$~K  and the small system sizes (see Table.\ref{tab:sim}).
The results seem consistent with the properties of ice mycrocristallites
(see Fig.\ref{bypii}). 

The (Basal) plane exhibits a divergence of $\Gamma(q)$ consistent
with a smooth interface.  A best fit to the model
of Eq.4 provides $w=6.5$ and
$g''=6.0\cdot 10^{15}$~J/m$^4$, while a constrained fit with $w=0$ yields
$g''=12\cdot 10^{15}$~J/m$^4$ and squared deviations that are one order
of magnitude larger.

For the (pII) plane we studied the spectrum along (pII)[Basal] and 
results are  suggestive of a rough
interface, with $\Gamma(q)$ apparently converging to $\Sigma$.
Unconstrained fits to the model of Eq.4 yield unphysical results,
with a negative $w$. A constrained fit with $w=0$
provides $g''=3.6\cdot 10^{15}$~J/m$^4$.

\begin{figure}[t]
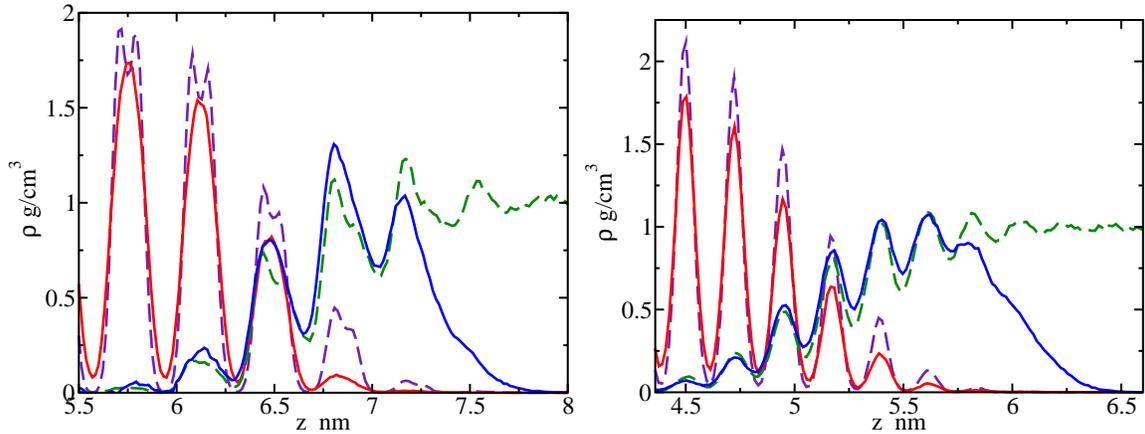

\centering
\includegraphics[clip,scale=0.27]{profile.comp.basal.pii.eps}
\includegraphics[clip,scale=0.27]{profile.comp.pii.pi.eps}
\caption{\label{prof.comp}
Density profiles of solid like (dashed line) and liquid like (full line)
ice/vapor (full lines) and ice/liquid (dashed lines) interfaces as
a function of perpendicular distance $z$ at $\Delta T=-2$~K. From left to right
results are shown for the (Basal) and (pII) planes. Rest of captions
as in Fig.3 of paper.
}

\end{figure}

As for the (pI) plane, the fluctuations of ice/film and film/vapor surfaces
are very well described by results from independent ice/water and
water/vapor interfaces. We checked that also the local
interfacial structure of the premelting film
 as described by the density profiles is similar to the
independent interfaces, as is seen in Fig.\ref{prof.comp}.

\subsection{Fitting procedure}

Fitting of $\Gamma(q)$ to the model of Eq.4 is difficult, because
$\Gamma_{\iqv}$ is typically orders of magnitude larger
than $\Gamma_{\iq}(q)$ and $\Gamma_{\qv}(q)$. For this reason we
find it more convenient to fit the results to $1/\Gamma(q)$. 
In this representation all the results are of similar
order of magnitude.
Fits were performed by simultaneously adjusting results for $\Gamma_{\iq}(q)^{-1}$,
$\Gamma_{\qv}(q)^{-1}$  and $\Gamma_{\iqv}(q)^{-1}$ with equal weights, from
the lowest wave-vector up to $q\approx 4.5$~nm$^{-1}$.
For rough interfaces the $\Gamma(q)^{-1}$ attain a constant value,
and for smooth interfaces they vanish as $q\to 0$. 

\subsection{System size effects}

\begin{figure}[t]
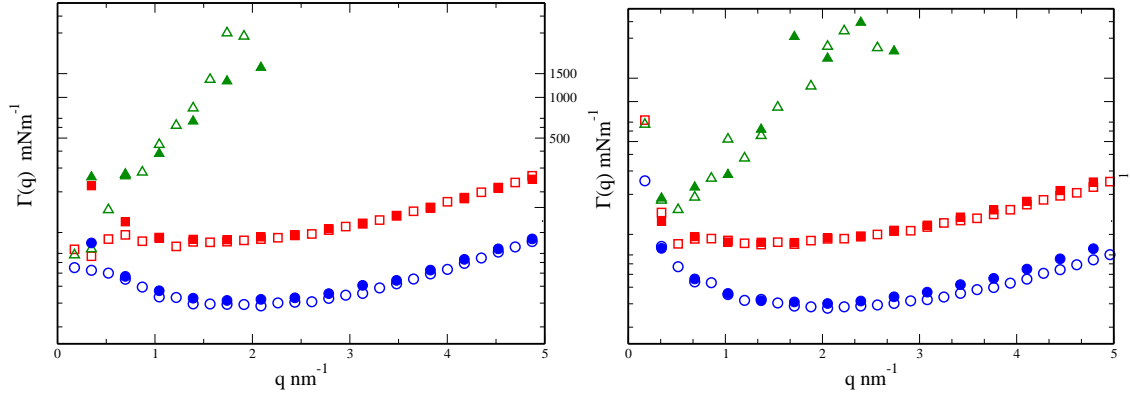

\centering
\includegraphics[clip,scale=0.30]{gamma_pi_basal_size.eps}
\includegraphics[clip,scale=0.30]{gamma_pi_pii_size.eps}
\caption{
System size dependence of $\Gamma(q)$ of the ice/vapor interface
for the (pI)[Basal] (left) and (pI)[pII] (right) systems studied in this work.
Results for the large system with $L_x\approx 36$~nm are shown with empty
symbols, and those for the small system with $L_x=18$~nm are shown with
filled symbols.
\label{systemsize}
}
\end{figure}

Since the decision as to whether a surface remains rough or smooth requires
to study the $q\to 0$ limit, it is important to assess whether the simulations
are subject to system size effects. In Fig.\ref{systemsize}, we compare $\Gamma(q)$ as
obtained for the ice/vapor interface for two different system sizes. The first system size is that
reported in the paper, with a long side $L_x$ of about 36~nm. The second is
one with equal dimensions for $L_y$ and $L_z$, but the long side $L_x$
half as large. For the (pI)[pII] system, the figure shows no significant system
size dependence (though the large system allows for twice as large density
of wave-vectors). For the (pI)[Basal] system, again the agreement between large
and small systems is very good, up to the lowest wave-vector. In this case we
find the largest system exhibits rough behavior, while the small one suggests
a smooth interface. We believe this discrepancy is not a system size effect,
but rather, a consequence of the neighborhood of the roughening transition and
the difficulty to control the temperature within an interval of about $0.5$~K. 

\section{Solution of the coupled Capillary-Wave + sine-Gordon model}

\subsection{Model and variational solution}

Consider a thin film sandwiched between a solid and a vapor phase. The state of
the film is described in terms of a solid/film profile,
$h_{\iq}(\rpar{}{})$, and a film/vapor profile, $h_{\qv}(\rpar{}{})$, 
which we denote for short as $h_1(\rpar{}{})$ and $h_2(\rpar{}{})$, 
respectively.
The solid/film interface
is described with the sine-Gordon model,\cite{chaikin95} and the 
film/vapor interface
with the Capillary Wave Hamiltonian in quadratic approximation.\cite{nelson04} 
Our coupled Hamiltonian for the quasi-liquid film is:
\begin{equation}\label{eq:Hsl}
 H_f =  \int d\rpar{}{} \left \{ \frac{1}{2}
\tilde\gamma_{iw} (\nabla h_{1})^2 -u \cos(k_z h_{1}) 
+ \frac{1}{2} \gamma_{wv} (\nabla h_{2})^2 +  \frac{1}{2}
g'' ( h_2 - h_1 )^2
\right  \}
\end{equation}
where $\tilde\gamma_{iw}$ is the stiffness of the ice/water interface, 
$u$ is a phenomenological coefficient dictating
the pining of the ice/film surface to discrete lattice spacings,
$k_z=2\pi/b$, with $b$ the inter-plane spacing in the direction
perpendicular to the interface. $\gamma_{wv}$ is the liquid/vapor surface
tension, and $g''$ is the second derivative of the interface potential
with respect to film thickness. Here it serves as a spring constant
for the harmonic fluctuations of the film thickness.\cite{dietrich88}

We seek solution of the partition function in terms of the
Fourier modes of both surfaces $h_{\alpha}(\qvec)$  ($\alpha$=1 or 2). Recognizing
that the second order approximation of the above result is quadratic
in the surface modes, we  consider a reference Hamiltonian of
independent harmonic oscillators with Gaussian statistics:
\begin{equation}
  H_0 = \frac{1}{2} \sum_{\qvec} {\bf h}(\qvec) {\bf G}^{-1}(\qvec) {\bf h}(\qvec)
\end{equation} 
where ${\bf h}(\qvec)=(h_1(\qvec),h_2(\qvec))$ and ${\bf G}^{-1}$
is  the covariance matrix:
\begin{equation}
    {\bf G}^{-1}(\qvec) = \left (
\begin{array}{cc}
    \sigma_{11}^2(\qvec) &  \sigma_{12}^2(\qvec) \\
    \sigma_{21}^2(\qvec) &  \sigma_{22}^2(\qvec) \\
\end{array}
\right )
\end{equation} 
with matrix components:
$\sigma^2_{11}(\qvec)=\langle h_1(\qvec)h^{*}_1(\qvec)\rangle$,
$\sigma^2_{22}(\qvec)=\langle h_2(\qvec)h^{*}_2(\qvec)\rangle$, 
$\sigma^2_{12}(\qvec)=\langle h_1(\qvec)h^{*}_2(\qvec)\rangle$
and $\sigma^2_{21}(\qvec)=\langle h_2(\qvec)h^{*}_1(\qvec)\rangle$.

The partition function of the reference Hamiltonian is:\cite{goldenfeld92}
\begin{equation}
 Z_0 = \int \prod_{\qvec} d {\bf h}(\qvec) 
  e^{-\frac{1}{2} \beta \sum_{\qvec} {\bf h}(\qvec)\, {\bf G}^{-1}(\qvec)\, {\bf
h}(\qvec)}
\end{equation} 
while the free energy is:
\begin{equation}
  F_0 = \frac{1}{2} k_B T \sum_{\qvec} \ln 
      \frac{ \operatorname{det} {\bf G}^{-1}(\qvec) }{(2\pi k_BT)^2}
\end{equation} 
This result can be expressed explicitly in terms of the elements of
the covariance matrix:
\begin{equation}
  F_0 = -\frac{1}{2} k_B T \sum_{\qvec} \ln 
  \left  [ (2\pi k_BT)^2 ( \sigma_{11}^2(\qvec) \sigma_{22}^2(\qvec) -
\sigma_{12}^4(\qvec) ) 
  \right ]
\end{equation} 

Next, we assess the free energy contribution from $H_f$ by
performing an average  over the Gaussian statistics
of the reference oscillators, $F_f=\prom{H_f}{0}$, 
with the result:
\begin{equation}
 F_f  =  \frac{1}{2} A \sum_{\bf q}
  \left \{ [g''+\tilde\gamma_{iw} q^2 ]\sigma_{11}^2(q)   + 
 [g''+\gamma_{wv} q^2 ]\sigma_{22}^2(q) -
 2 g'' \sigma_{12}^2(q) - u e^{-\frac{1}{2} k_z^2 \sum_{\qvec} \sigma_{11}^2(q)}
        \right \}
\end{equation}
where $q$ is the norm of ${\bf q}$. Notice we have
taken into account that the Gaussian average of
$\cos(k_z h_1(\rpar{}{}))$ is $\exp(-\frac{1}{2} k_z^2 \langle h_1^2(\rpar{}{})\rangle)$,
and further 
used Parseval's theorem to transform
averages of $h_1^2(\rpar{}{})$ and $h_2^2(\rpar{}{})$ into averages of their Fourier
components.

The total free energy is $F_0 + F_f$, and we obtain a solution in closed
form by seeking for the variational parameters $\sigma_{11}^2(q)$,
$\sigma_{22}^2(q)$ and $\sigma_{12}^2(q)$. The result of this
minimization yields:
\begin{equation}\label{eq:h2surf}
\begin{array}{ccc}
 \sigma_{11}^2(q)  & = &
 \displaystyle{     \frac{k_BT}{A} \frac{g'' + \gamma_{wv} q^2}{
[ w + g'' + \tilde\gamma_{iw} q^2 ][  g'' + \gamma_{wv} q^2 ] - g''^2} 
} \\
 & & \\
 \sigma_{22}^2(q) & = &
  \displaystyle{      \frac{k_BT}{A} \frac{w + g'' + \tilde\gamma_{iw}
q^2}{
[ w + g'' + \tilde\gamma_{iw} q^2 ][  g'' + \gamma_{wv} q^2 ] - g''^2}  
} \\
& & \\
 \sigma_{12}^2(q) & = &
   \displaystyle{      \frac{k_BT}{A} \frac{g''}{
[ w + g'' + \tilde\gamma_{iw} q^2 ][  g'' + \gamma_{wv} q^2 ] - g''^2}  
}
\end{array}
\end{equation}
where the roughness parameter $w$ is 
\begin{equation}\label{eq:wcondition}
  w = u k_z^2 e^{-\frac{1}{2} k_z \sum_q \sigma_{11}^2(q)}
\end{equation}

To solve for this self consistent condition, we
ignore surface anisotropy, which is small for ice (see next section),
and approximate:
\begin{equation}
   \sum_{\bf q} \sigma_{11}^2(q) = \frac{k_B T}{2\pi} 
  \int_0^{q_{max}} q d q
 \frac{g'' + \gamma_{wv} q^2}
     {w g'' + (g''\Sigma + w \gamma_{wv})q^2 + \gamma^2 q^4} \,
\end{equation} 
where $q_{max}$ is an ultra-violet cut--off
and we have introduced $\Sigma=\tilde\gamma_{iw}+\gamma_{wv}$ and
$\gamma^2 = \tilde\gamma_{iw}\gamma_{wv}$ for short. 

This integral may be solved in real space
along the lines indicated in Ref.\cite{macdowell15},
with the result:
\begin{equation}\label{eq:roughnes}
 \sum_q \sigma_{11}^2(q) = C \ln ( 1 + \frac{pr}{wg''}\,q_{max}^2 )
+ D \ln ( 1 + \frac{\gamma^2}{pr}\,q_{max}^2 )
\end{equation} 
where:
\begin{equation}
 p = g''\Sigma + w\gamma_{wv}
\end{equation} 
\begin{equation}\label{eq:rsq}
  r = \frac{p + (p^2-4w g''\gamma^2)^{1/2}}{2 p}
\end{equation} 
\begin{equation}
  C = \frac{k_BT}{4\pi}\,\frac{g'' r p - g''\gamma_{wv} w}{r^2 p^2 - w g''\gamma^2} 
\end{equation} 
\begin{equation}
  D = \frac{k_BT}{4\pi}\,\frac{p r}{\gamma^2}\frac{\gamma_{wv} r p - g''\gamma^2}{r^2 p^2 - w g''\gamma^2} 
\end{equation} 
For H$_2$O, where  $\gamma_{wv}>\tilde\gamma_{iw}$ at the triple point, 
$r$ is always  positive and
real. 

As we shall see, the outcome of the self consistent condition
is mainly dictated by the behavior of \Eq{roughnes} at small $w$, whence,
we assume $p^2>>4w g''\gamma^2$, so that expanding the
square root in \Eq{rsq} to zero order, the coefficients $C$ and $D$ simplify to:
\begin{equation}\label{eq:c0}
C = \frac{k_BT}{4\pi}\frac{g''^2\Sigma}{(g''\Sigma + w \gamma_{wv})^2} 
\end{equation} 
and
\begin{equation}\label{eq:d0}
  D = \frac{k_BT}{4\pi}\frac{\gamma_{wv}}{\tilde\gamma_{iw}}\frac{g'' +
w}{g''\Sigma+w\gamma_{wv}} 
\end{equation} 
By use of \Eq{wcondition}, \Eq{roughnes} and Eq.\ref{eq:c0}-\ref{eq:d0}, 
we obtain the self-consistent condition in closed form as:
\begin{equation}\label{eq:wcondtot}
 w = u k_z^2 
 \left ( 1 + \frac{g''\Sigma + w\gamma_{wv}}{w g''}\, q_{max}^2  \right)^{-\mu} 
     \cdot 
 \left ( 1 + \frac{\gamma^2}{g''\Sigma + w \gamma_{wv}}\, q_{max}^2 \right)^{-\tau} 
\end{equation} 
where $\mu=\frac{1}{2} k_z^2 C$ and $\tau = \frac{1}{2} k_z^2 D$. 

\begin{figure}[t]
\centering
\includegraphics[clip,scale=0.30]{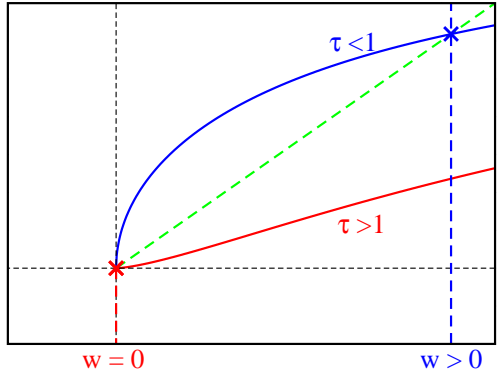}
\caption{
Sketch of the solution of $w \propto y(w)$  dictating the roughness
of a solid/liquid interface. The green line is a straight line of
unit slope. Blue and red lines correspond to y(w) (\Eq{yfunc}) for $\tau>1$ and
$\tau<1$, respectively. The equilibrium roughnes $w$ is given by the
intersection of either curve with the straight green line.
For $\tau>1$, only one root at $w=0$
exists (rough interface). For $\tau<1$, an additional route
$w>0$ is found (smooth interface).
\label{regimen_tau}
}
\end{figure}

In order to understand the significance of this result, let us first consider
the case of the ice/water interface (film of infinite thickness), 
such that $g''\equiv 0$. In
that case, $\mu\equiv 0$, and we get:
\begin{equation}\label{eq:condiw}
 w = u k_z^2 \left ( 1 + \frac{\tilde\gamma_{iw}}{w}\, q_{max}^2 \right)^{-\tau_{iw}}
\end{equation} 
with $\tau_{iw}= \frac{\pi k_BT}{2\tilde\gamma_{iw} b^2}$. It follows that $w$
is the real root of the auxiliary equation $x=y(x)$, with
\begin{equation}\label{eq:yfunc}
  y(x) =  \left ( 1 + \frac{a}{x} \right )^{-\tau_{iw}}
\end{equation} 
and $a$ a constant. i.e., $w$ is
given by the intersection of a straight line of unit slope
with the auxiliary function $y(x)$.
This function goes through the origin with positive slope, then
bends and eventually becomes constant. For $\tau_{iw}<1$, $y(x)$ meets the
origin with infinite slope, and must therefore cross the straight line 
of unit slope at finite $x$. If, on the other hand, $\tau_{iw}>1$, $y(x)$ meets
the origin with zero slope and there is only one root at $x=0$
(see Fig.\ref{regimen_tau}). Whence,
we encounter a roughening transition from a smooth surface
(finite $w$) to a rough surface ($w=0$) at $\tau_{iw}=1$.\cite{safran94}
This corresponds exactly to the  roughening transition of the sine-Gordon
model.\cite{chaikin95}

Let us now consider the general case
of an ice surface covered by a quasi-liquid film of finite depth,
with finite $g''$. The solutions for $w$ are now given by \Eq{wcondtot}, 
which, in the limit $w\ll g''$ becomes:
\begin{equation}
 w = u k_z^2 
 \left ( 1 + \frac{\Sigma}{w}\, q_{max}^2 \right)^{-\mu_{\iq}} 
     \cdot 
 \left ( 1 + \frac{\gamma^2}{g''\Sigma}\, q_{max}^2 \right)^{-\tau_{if}} 
\end{equation} 
where $\mu_{\iq} = \frac{\tilde\gamma_{iw}}{\Sigma} \cdot \tau_{iw}$ and
$\tau_{if} = \frac{\gamma_{wv}}{\Sigma} \cdot \tau_{iw}$.
This result shows the dramatic consequences of pining the liquid film
on the ice surface. Particularly, the second parenthesis of
the right hand side, which drives the roughening transition 
for the case $g''\equiv 0$, is now a constant. On the other hand,
the first parenthesis, which was unity in that case, becomes
responsible for driving the roughening transition.
In fact, it  takes exactly the same form as \Eq{condiw},
albeit, with a completely
different exponent, which differs from $\tau_{iw}$ by a factor
$\tilde\gamma_{iw}/\Sigma$. For our model,
 this is about 1/3 at the triple point. 
The implication is that under conditions where the ice/water surface
has become rough, with $\tau_{iw} \geq 1$, we expect  that 
$\mu_{\iq}$ will be 
largely below unity. Hence, the coupling of the film/vapor interface
to the ice/film surface via $g''$ drives the rough ice/water interface
into a smooth surface. In practice, the effect is significant for large
$g''$. For finite but small $g''$, the root occurs extremely close
to $w=0$, and the smoothening is then only apparent at very large
length-scales. Our results are consistent with previous studies showing
the sensitivity of roughening to monolayer surface adsorption.\cite{akutsu01}

\subsection{Roughening anisotropy}

\label{roughan}

Ordinary ice has hexagonal symmetry and is therefore not strictly isotropic.
Particularly, for the ice/water interface at $\Delta T=-2$~K, the 
(pI) crystal plane has slightly different principal 
stiffness coefficients for the [Basal] (call it $x$) and [pII] (call it $y$) directions (i.e.,
$\tilde\gamma_{iw}(x)=26.14$~mN/m and $\tilde\gamma_{iw}(y)=27.18$~mn/m, respectively).
To account for this anisotropy, the Hamiltonian of \Eq{Hsl} should be 
replaced by:
\begin{equation}\label{eq:Hsla}
 H_f =  \int d\rpar{}{} \left \{ \frac{1}{2}
\tilde\gamma_{iw}(x) \derpar{h_{1}}{x}{}^2 + \frac{1}{2} \tilde\gamma_{iw}(y)
\derpar{h_{1}}{y}{}^2 -u \cos(k_z h_{1}) 
+ \frac{1}{2} \gamma_{wv} (\nabla h_{2})^2 +  \frac{1}{2} g'' ( h_2 - h_1 )^2
\right  \} 
\end{equation}
The solution of this Hamiltonian in Fourier modes is exactly as for the isotropic
case, provided one replaces $\tilde\gamma_{iw} q^2$ 
by $\tilde\gamma_{iw}(x) q_x^2 + \tilde\gamma_{iw}(y) q_y^2$.\cite{fisher83}
Particularly, the result for $\sigma_{11}^2$ in \Eq{h2surf}
should be replaced by:
\begin{equation}\label{eq:h2surfany}
\begin{array}{ccc}
 \sigma_{11}^2(\qvec)  & = &
 \displaystyle{     \frac{k_BT}{A} \frac{g'' + \gamma_{wv} q^2}{
[ w + g'' + \tilde\gamma_{iw}(x) q_x^2 + \tilde\gamma_{iw}(y) q_y^2 ][  g'' + \gamma_{wv} q^2 ] - g''^2} 
} 
\end{array}
\end{equation}
For large but finite surfaces of circular or square shape the sum 
$\sum_{\qvec} \sigma_{11}^2(\qvec)$ required to measure $w$ in \Eq{wcondition}
is effectively performed from a low wave-vector
$\qvec=(\frac{2\pi}{L},\frac{2\pi}{L})\to (0,0)$,
and the anisotropy of the stiffness coefficients is
inconsequential.\cite{akutsu15} 

In our systems, however, we study surfaces that are very thin
in one direction and large in the other. Accordingly, the large wave-length
modes in the short direction are cut-off and the lower cut-off
is anisotropic, since now $\qvec=(\frac{2\pi}{L_x},\frac{2\pi}{L_y})$
and $L_x\ne L_y$. Whence, for the (pI)[Basal] direction
we have an effective lower cut-off at
$\qvec=(0,\frac{2\pi}{L_y})$, while in the (pI)[pII] direction
the lower-cutoff is $\qvec=(\frac{2\pi}{L_x},0)$.
As a result, the sum $\sum_{\qvec} \sigma_{11}^2(\qvec)$ now
depends on the system geometry. Accordingly, the self-consistent
condition for the roughness parameter, \Eq{wcondition}, also becomes
anisotropic. 

The effect is weak, however, because the anisotropy in the
stiffness coefficient is small.
Indeed, for the crystal/melt interface of NaCl we
have  checked previously that for systems well above the
roughening transition, the results for an elongated
system are fully equivalent to those of a square interface.\cite{benet15}

In our case the problem is more subtle, because we are close to the roughening
transition.  Unfortunately, this effect  cannot be 
described analytically, because the integral of \Eq{h2surfany} can
be solved in circular coordinates, but not in rectangular coordinates 
as required here.\cite{akutsu15}

Nevertheless, assuming we can obtain the root of $w$ using
the exponent $\mu_{if}$ for the isotropic case by merely replacing
$\Sigma=\gamma_{lv}+\tilde\gamma_{iw}$ with  the corresponding anisotropic
stiffness,
we find the exponent for roughening in the (pI)[Basal] direction
is $\approx (72+27)/(72+26)=1.01$ times larger than that for the (pI)[pII] direction,
whence, the (pI)[Basal] system may become effectively rough
at a temperature where (pI)[pII] remains smooth, as observed in
our simulations.

\subsection{Asymptotic behavior of the surface fluctuations}

\begin{figure}[t]
\centering
\includegraphics[clip,scale=0.30]{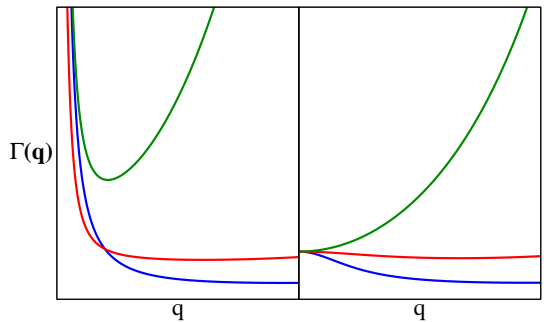}
\caption{
Results for $\Gamma(q)$ as obtained from the model of \Eq{h2surf}
for the ice/film (blue), film/vapor (red) and coupled fluctuations
of the ice/film and film/vapor surfaces (green).
Left: Behavior expected for a smooth interface, with
finite  $g''$ and $w$.  Right: Behavior expected for
a rough interface, with finite $g''$ but  $w=0$. 
\label{rvs}
}
\end{figure}

The behavior of the surface Fourier modes adopts distinct behavior
in the limit $q\to 0$, depending on whether $g''$ and $w$ are null
or finite (Fig.\ref{rvs}). 

This allows to tell whether the surface is rough ($w=0$)
or smooth $w>0$ from the behavior of the spectrum of fluctuations.

Defining:
\begin{equation}
 \begin{array}{ccc}
   \Gamma_{\iq}(q) & = & \frac{k_BT}{A \prom{h_{\iq}(q)h^{*}_{\iq}(q)}{} q^2} \\
   \Gamma_{\qv}(q) & = & \frac{k_BT}{A \prom{h_{\qv}(q)h^{*}_{\qv}(q)}{} q^2} \\
   \Gamma_{\iqv}(q) & = & \frac{k_BT}{A \prom{h_{\iq}(q)h^{*}_{\qv}(q)}{} q^2} \\
 \end{array}
\end{equation} 
we find, from \Eq{h2surf}:

\begin{description}
\item Limit of $q\to \infty$.

For sufficiently large $q$, both surfaces of the quasi-liquid layer
 behave like rough and independent  ice/water and water/vapor interfaces:
\begin{equation}
 \begin{array}{ccc}
   \Gamma_{\iq}(q) & \to & \tilde\gamma_{iw} \\
   \Gamma_{\qv}(q) & \to & \gamma_{wv} \\
   \Gamma_{\iqv}(q) & \to & \frac{q^2}{g''} \\
 \end{array}
\end{equation}

\item Limit of  $q\to 0$.

In the limit of small $q$, the behavior is distinctly different depending
on whether the ice/vapor interface is smooth ($w>0$) or rough ($w=0$):

\begin{description}
\item Smooth interface ($w>0$).

For a smooth interface, the fluctuations remain finite at zero wave-vector,
and the $\Gamma_{\alpha\beta}$ diverge.
\begin{equation}
 \begin{array}{ccc}
   \Gamma_{\iq}(q) & \to & \frac{w}{q^2} \\
   \Gamma_{\qv}(q) & \to & \frac{w g''}{(g''+w) q^2} \\
   \Gamma_{\iqv}(q) & \to & \frac{w}{q^2} \\
 \end{array}
\end{equation} 

\item Rough interface ($w=0$).

For a rough interface, the fluctuations diverge and the interface
behaves globally as a rough surface with stiffness
$\Sigma=\tilde\gamma_{iw}+\gamma_{wv}$.
\begin{equation}
 \begin{array}{ccc}
   \Gamma_{\iq}(q) & \to & \Sigma \\
   \Gamma_{\qv}(q) & \to & \Sigma \\
   \Gamma_{\iqv}(q) & \to & \Sigma \\
 \end{array}
\end{equation}
\end{description}
\end{description}

Notice that the latter behavior is also observed for a smooth
interface in the
range of wave-vectors
$g'' w\ll g'' \Sigma q^2 \ll \gamma^2 q^4$ 
provided $w\ll g''$. Therefore, for very small $w$ it is
required to attain a regime of very small $q$ to discriminate
between a rough and a smooth surface.